# Allocative efficiency in public research funding: can bibliometrics help?[†][‡]

*Giovanni Abramo[a,b,][*], Ciriaco Andrea D'Angelo[a], Alessandro Caprasecca[a]*

[a] Laboratory for Studies of Research and Technology Transfer
Department of Management, School of Engineering
University of Rome "Tor Vergata"

[b] Italian Research Council

**Abstract**

The use of outcome control modes of research evaluation exercises is ever more frequent. They are conceived as tools to stimulate increased levels of research productivity, and to guide choices in allocating components of government research budgets for publicly funded institutions. There are several contributions in the literature that compare the different methodological approaches that policy makers could adopt for these exercises, however the comparisons are limited to only a few disciplines. This work, examining the case of the whole of the "hard sciences" of the Italian academic system, makes a comparison between results obtained from peer review type of evaluations (as adopted by the Ministry of Universities and Research) and those possible from a bibliometric approach (as developed by the authors). The aim is to understand to what extent bibliometric methodology, which is noted as relatively inexpensive, time-saving and exhaustive, can complement and integrate peer review methodology in research evaluation.

[†] Abramo, G., D'Angelo, C.A., Caprasecca, A. (2009). Allocative efficiency in public research funding: can bibliometrics help? *Research Policy*, 38(1), 206-215. DOI: 10.1016/j.respol.2008.11.001
[‡] The authors wish to thank two anonymous referees for their valuable comments and suggestions.
[*] Corresponding author, Dipartimento di Ingegneria dell'Impresa, Università degli Studi di Roma "Tor Vergata", Via del Politecnico 1, 00133 Rome, ITALY. Tel +39 06 72597362, Fax +39 06 72597305, abramo@disp.uniroma2.it.

# 1. Introduction

In the present knowledge-based economy, the governments of the major industrialized nations are under strong pressure to render public scientific infrastructure ever more effective in sustaining the competitiveness of domestic industry. The rising costs of research and the tight restrictions in public budgets, call for the adoption of more efficient systems of resource allocation. To this end, many countries have begun to impose national exercises in research evaluation. The objectives of these are two-fold: to aid in allocating resources according to merit and to stimulate increased levels of research productivity on the part of the funding recipients. In Europe, the experiences of Great Britain's Research Assessment Exercise (RAE), as well as the analogous Triennial Research Evaluation in Italy (termed the "VTR"), and the Netherlands Observatory of Science and Technology offer explicit examples of national exercises with either or both objectives. Outside Europe there are the corresponding experiences of the Australian Research Quality Framework and the New Zealand Performance-Based Research Fund.

In Italy, the evaluation question has been the subject of a lively debate that sees opposition between public research institutions on the one hand, lamenting chronic deficits in the resources they need to fulfill their role, and the national government on the other, impelled towards stringent policies by the size of the deficit and the necessity of controlling expenses.

In general, the approaches currently in use for research evaluation can be assigned to two categories, either peer review or bibliometric techniques. In peer review, judgment is entrusted to a panel of experts that synthesizes and assigns a judgment based on the examination and appraisal of certain parameters, such as relevance, originality, quality, or potential socio-economic impact of research outputs. Bibliometric techniques are based on indicators, elaborated from the data which can be found in specialized publications databases, concerning the quantity of research output that is codified as scientific articles, the quality of the articles (in terms of number of citations received) and/or the quality of the journals in which they are published (their "impact factor").

The peer review approach is certainly the more widespread, though not free of limitations. These limitations can be traced, in large part, to certain forms of subjectivity arising in the assessments (Moxham and Anderson, 1992; Horrobin, 1990): the selection of the experts called on to evaluate the outputs and the process of assessing levels of quality by these experts. The evaluation outcomes will not be insensitive to the model to be applied in giving the evaluations, nor to the selection of a measurement "scale" for the research outputs. The significant times and costs for peer types of evaluation exercises present another sensitive point, to the extent that peer evaluations can generally only be restricted to samples of the production of research institutions: they do not adapt to measuring complete levels of productivity.

The bibliometric approach, in contrast, offers clear advantages. Bibliometric assessments are economical, non invasive, and simple to implement; they permit updates and rapid inter-temporal comparisons; they are based on more objective[3]

---

[3] Weingart (2003), observed that bibliometric indicators themselves are indirectly based on peer evaluation, since articles published on refereed journals must have gone through an evaluation process, based in part on peer decisions. The authors note, though, that citations counts reflect a much wider evaluation than that of a referee panel.

qualitative[4]-quantitative data and are capable of examining a higher representation of the universe under investigation and adapt well to international comparisons. Bibliometric methods also have their own risks that can influence validity (Van Raan, 2005a; Georghiou and Larédo, 2005). There exist intrinsic limitations in the ability of the journal listings to serve as representative of the editorial universe, during bibliographic evaluations, and to be equally representative of the disciplinary sectors under consideration. The publications in international journals that are subject to bibliometric evaluations are generally highly representative of research output in hard sciences, but not at all of research in arts and humanities. Errors can be made in attributing publications and their relative citations to the source institutions of the authors (Moed, 2002). The difficulty of giving due consideration to differences in the level of productivity between disciplines also determines significant distortions in comparisons at the aggregate level, when comparisons are made on aggregates among institutions representing a variety of different disciplinary areas (Abramo and D'Angelo, 2007). Finally, the most obstinate obstacle confronting the bibliometrician is the accurate attribution of articles to their true authors and institutions, due to homonyms in author names and to the failure of the source databases to indicate the individual connections between the authors listed and the accompanying lists of research institutions. This obstacle has, until now, severely limited the full use of bibliometric evaluations at a national scale. The authors have overcome this barrier by bringing into service, at the Italian level, a disambiguation algorithm that permits the unequivocal attribution of the publication to its author, with a margin of error of less than 2%.

This recent advance now makes it possible to integrate the two approaches and permit:
- a wider use of bibliometrics for evaluation of quality and efficiency of research activities in those areas of scientific investigation that are well represented by publication in international journals (i.e. hard sciences);
- the use of peer review for evaluation of outputs where international scientific publications do not provide a reliable proxy of research output (in socio-economic areas and especially in the arts and humanities), or for evaluation of aspects other than output quality, such as socio-economic impact.

This study has the specific objective of investigating to what extent bibliometric methodology, being relatively inexpensive, time saving and exhaustive, can complement and integrate peer review.

The study will examine the Italian case and in particular the "hard science" disciplines[5] and then proceed with a comparison of results from the first national peer review exercise, the VTR, completed in 2006, and the rating of the same universities elaborated using bibliometric methodology.

A particular intention of the study is to investigate the presence of potential elements of replication and/or complementarities in the results of the peer review and bibliometric methods. To do this, the comparison will include correlation analysis of the VTR peer evaluation results with bibliometric measures that can be readily made from the same outputs.

---

[4] By "qualitative" data we mean articles weighed by an index of quality (meaning citations or journal impact factor).

[5] According to the Italian university order, "hard sciences" are classified into 9 disciplinary areas: mathematics and computer sciences; physics; chemistry; earth sciences; biology; medicine; agriculture and veterinary sciences; industrial and information engineering.; civil engineering and architecture. Of these, civil engineering and architecture was discarded from consideration because the SCI™, our source of bibliometric data, does not cover the full range of research output.

One should note that, if the scope of the policy maker is to allocate resources in function of the scientific excellence of research institutions, then the method of selecting the outputs to represent these institutions in evaluation is a critical area of examination. For this, a second intention of this study is to evaluate, again by reference to the VTR, the effectiveness of the process of selecting outputs for submission, as practiced by the universities at the outset of evaluation.

A third concern is the question of whether the quality dimension alone of research output is a sufficient indicator of scientific excellence, or whether both quality and productivity dimensions should be considered. Since it may be that the quality of a limited sample of outputs does not necessarily reflect the average "qualitative productivity[6]" of a research institution, we will determine the difference between the ratings derived from the VTR (indicating the quality of research outputs) and the contrasting ratings of the research productivity of the institution, as determined on a bibliometric basis.

The study is organized as follows: the following section gives a summary of previous contributions to the literature concerning methodologies and results of the methods used for evaluation. Section 3 describes the peer review analytical model adopted by the Italian VTR. Section 4 provides a summary description of the specific bibliometric analysis methodology developed for the scope of this work: the field of observation, sources and indicators elaborated. Section 5 illustrates the results of the analysis with reference to each of the above research questions that motivated the study. The paper concludes with a summary of the study's salient features and the authors' comments.

## 2. Assessment methods in academia: evidence and open questions

Numerous studies have highlighted the fundamental roles of evaluations, both as incentives to scientific excellence and as instruments for strategic choices on the part of political decision-makers (Van Raan, 2005b; Narin and Hamilton, 1996; Moed et al., 1995). One of the questions still open to major debate among scholars is that concerning the choice of the most satisfactory evaluation system to be used. Science and Public Policy (2007) published a special issue dealing with academic opinion on directions for science policy and research evaluation; this issue illustrated the dichotomies separating metrics and peer review It is clear that in current practice the peer review approach is certainly the one in greatest use. It is held as being well adapted to determining the quality of research, thanks to its direct observation of a number of factors, including even those such as the composition of research teams, the industrial significance and potential socio-economic impact of the results (Reale et al., 2006).

On the other hand, some authors demonstrate that bibliometric methods, particularly those using multiple indicators (Martin, 1996), provide useful information that could counterbalance shortcomings and mistakes in peer judgments (Aksnes and Taxt, 2004; Rinia et al., 1998). Peer evaluation is potentially affected by built-in distortions from subjectivity in assessments (Moxham and Anderson, 1992; Horrobin, 1990): these are distortions that can occur at several levels, including the phase of using peer judgment to gauge output quality, but also in the earlier steps of selecting outputs to be submitted for evaluation and selecting the experts that will carry out the assessments.

---
[6] The definition of qualitative productivity is provided in section 4.

Subjective evaluations can also be affected by real or potential conflicts of interest; from the inclination to give more positive evaluations to outputs from known and famous researchers than to those of the younger and less established researchers in the sector; or from failure to recognize qualitative aspects of the product (a tendency that increases with increasing specialization and sophistication in the work). In addition, peer methodology does not have universal consistency, since mechanisms for assigning merit are established autonomously by the various evaluation committees, thus exposing comparisons linked with this methodology to potential distortions.

Peer methodology has certainly shown itself to be highly flexible, adapted as it is to evaluation of all potential forms of research output. Since it can observe a number of dimensions, it is also particularly effective where the intention is to surpass analysis of scientific quality of outputs and investigate further complimentary details, such as the role of the source organization as a pole of attraction, or its level of international activity, or the potential socio-economic impact of the outputs.

Peer review methods, however, do involve notable costs and long periods for completion. Take the case of the Italian VTR, which examined a sample of approximately 14% of the product realized by the Italian academic system for the period under observation: the evaluation featured direct costs of 3.5 million euros and more than 12 months for completion. The RAE, in England, which deals with approximately half of the total portfolio of outputs of the institutions evaluated, costs roughly 20 million euros per exercise. One must add to these accounting costs the opportunity costs of foregone research by the scientific personnel of the institutions surveyed, who devote time and efforts in selecting outputs to participate in the evaluation.

Bibliometric methods, on the other hand, prove to be more efficient at guaranteeing lower direct costs and present a total absence of opportunity costs. With these methods, performance is measured by means of qualitative-quantitative data that can be extracted from specialized databases, without need of "in the field" data collection, guaranteeing notable savings in time. Furthermore, the simplicity of the analytical processes consents rapid updating and inter-temporal comparison, while the level to which the full universe is investigated, plus the potential to proceed to international comparisons, constitute other strong points of this method.

There are certainly more than just a few authors who recommend caution when using bibliometric indicators in benchmarking universities and research institutions. Such approaches are the object of several potential methodological critiques. An initial problem concerns the lack of homogeneity in the coverage of various sectors of scientific investigation, on the part of the starting databases (Van Raan, 2005a). Several authors also describe the limitations inherent in considering the citations received by an article, or the impact factor of its relative journal of publication, as indicators of the actual quality of the article. These limitations can become more insidious if aggregate evaluations are then made, in particular if there is a failure to consider variations in the habits of publication and citation which occur between the different research sectors (Korevaar and Moed, 1996). These limitations, however, can be overcome, as demonstrated by the authors (Abramo and D'Angelo, 2007), when one initiates with a bottom-up approach: first identifying scientific production on the basis of author name, then disambiguating the authors and organizations responsible for the publication, then continuing from the output of individual researchers to aggregate successively the outputs by sector and (with appropriate normalization) by disciplinary area and university.

The literature offers various contributions dedicated to the comparison of peer and bibliometric methodologies, typically based on correlation analysis of actual assessments obtained from these two methods. The correlations result as being significant, though not perfect. Rinia et al (1998), for example, studied the links between results from peer evaluations conducted in 1996 by FOM (the Foundation for Fundamental Research on Matter) on material physics research units in Dutch universities, and a set of bibliometric indicators calculated for the period 1985 to 1994 on the output of the same groups, obtaining a positive and significant correlation between the two methods. Oppenheim (1997), suggests that citation counting provides a reliable indicator of research performance, after observing in a study carried out between 1988 and 1992 on articles in anatomy, genetics and archeology, that there was a correlation between citation counts and the UK Research Assessment Exercise (RAE) score. In 2003, Oppenheim and Norris compared outcomes of a citation study to RAE, for academic staff involved in archaeology and confirmed the presence of a statistically highly significant correlation.

Aksnes and Taxt (2004) compared results from a 1997 to 2002 peer evaluation at the University of Bergen, for research areas in departments of chemistry, mathematics, informatics, biology, and earth sciences, with results from bibliometric indicators for the same departments as elaborated from the 1995 to 2000 Thomson Reuters ICR (Institutional Citation Report): the analysis revealed a positive and significant correlation, though not strong. Finally, Reale et al. (2006), analyzed four of the disciplinary areas assessed in Italy's VTR exercise (chemistry, biology, humanities and economics), correlating peer assessments of articles in these disciplines with corresponding measures of the impact factor of the journals that published the articles: the correlations were positive, but not strong.

The preceding studies, however, suffer from limitations, especially in their methodology. In particular, the comparisons draw on fields of observation limited to only a few disciplinary sectors or to just one sector (Rinia et al., 1998; Reale et al., 2006), or even to a single institution (Aksnes and Taxt, 2004). In some cases, due to lack of full availability, the studies also compare data sets gathered in different time periods, and thus suffer still further.

In contrast to the cases of the previous publications, the scope of comparative investigations proposed for the current study is characterized primarily by an exhaustive field of observation, covering research in the hard sciences (fields which certainly do not lack for bibliometric significance[7]) as produced by the national totality of Italian universities.

## 3. The VTR peer evaluation

In December 2003, the Italian Ministry for Universities and Research (MUR) launched its first-ever Triennial Research Evaluation (VTR), which for this opening occasion referred to the period 2001-2003. The national Directory Committee for the Evaluation of Research (CIVR) was made responsible for conducting the VTR. The assessment system would be designed to evaluate research and development carried out by public research organizations (102 in total), including both universities and research organizations with MUR funding[8].

---

[7] As Van Raan (2005b) states "If international journals are the dominating or at least a major means of communication in a field, then in most cases bibliometric analysis is applicable".

As a first step, the CIVR selected experts for 20 panels, one for each disciplinary area. Universities and public research organizations were then asked to autonomously and submit research outputs to the panels[9]: outputs were not to be from more than 50% of the full-time-equivalent researchers working in the institution in the period under observation[10]. Outputs acceptable were limited to articles, books, and book chapters; proceedings of national and international congresses; patents and designs; performances, exhibitions and art works. Thus the VTR was designed as an ex-post evaluation exercise focused on the best outputs produced by Italian research institutions.

In the next step, the panels assessed the research outputs and attributed a final judgment to each product, giving ratings of either "excellent", "good", "acceptable" or "limited". The panels were composed of 183 high level peers appointed by the CIVR, and called on additional support from outside experts. The judgments were made on the basis of various criteria, such as quality, relevance and originality, international scope, and potential to support competition at an international level. To this purpose, the following quality index ($R_{ji}$) was used for ranking research institution "j" in area "i":

$$R_{ji} = \frac{1}{T_{ji}} \times [E_{ji} + 0.8\, G_{ji} + 0.6\, A_{ji} + 0.2\, L_{ji}] \qquad [1]$$

Where:

$E_{ji}$; $G_{ji}$; $A_{ji}$; $L_{ji}$ = numbers of "excellent, good, acceptable" and "limited" outputs submitted by the j$^{th}$ research institution in area i

$T_{ji}$ = total number of outputs submitted by the j$^{th}$ research institution in area i

A final report ranks institutions based on their results under the quality assessment index[11]. The rankings were realized at the level of single disciplinary areas, but not at the level of aggregates. The universities within each area were subdivided by dimension of their size, in function of the number of outputs submitted. As an example, Table 1 shows the ranking list of Italian "large" universities based on $R_{ji}$, in the disciplinary area "mathematics and computer science". Table 1, in addition to the dimensional ranking, gives the ranking within the universe of institutions active in the area under examination. Table 2 presents the example of the specific ratings obtained by the University of Rome "Tor Vergata", in the 11 disciplinary areas for which it submitted outputs.

---

[8] The remainder of this text, however, makes explicit reference only to universities.
[9] They were also asked to provide the CIVR with sets of input and output data for the institution and for the individual scientific areas within it.
[10] According to some authors, the type of procedure adopted by the CIVR would permit an evaluation exercise to overcome the inflationary effects of the "publish or perish" paradigm (Viale and Leydesdorff, 2003).
[11] In point of fact, the VTR ended with a third phase, in which CIVR integrated the outcome of the panels' analysis with its own analysis of data and information collected by institution and scientific area. For the purposes of the paper, this phase will not be considered (the results are available at http://vt2006.cineca.it/).

[Table 1]
[Table 2]

The magnitude of the VTR effort can be suggested by a few pertinent facts: the evaluation included 102 research institutions (77 universities and 25 public research organizations) and examined about 18,000 outputs, drawing on 20 "peer" panels (14 for disciplinary areas[12] and 6 for interdisciplinary sectors[13]), 183 panelists and 6,661 reviewers, with the work taking 12 months, and with financial costs mounting to 3.5 million euros.

## 4. The ORP bibliometric evaluations

Data used for the bibliometric analysis of the Italian academic system were extracted by the ORP (Observatory of Public Research for Italy) database. The ORP collects and sorts information on the scientific production by scientists in Italian public research laboratories, in a form that enables aggregation operations at higher levels (scientific sector, disciplinary area, faculty or school, university). The database was developed by extracting all publications (articles and reviews) listed in the Thomson Reuters Science Citation Index™ (Cd-Rom version) having at least one Italian affiliation[14]. The details of the specific methodologies used to construct the ORP are described in Abramo et al. (2008). The publications analyzed, as was the case for the VTR, refer to the period 2001 to 2003. In this period the number of scientists in Italian research institutions totaled 55,000, with 36,000 of these in the hard sciences. For the same period, the ORP indexes 71,047 publications, including the 7,513 research articles which were selected for VTR in the 8 disciplinary areas considered.

The normalized value[15] of the impact factor of the publishing journal (IF), at the time of the article was published, was considered as a proxy of the quality of the articles[16]. The authors are clearly aware of potential criticisms of the use of the impact factor of the journals as a proxy of the quality of each article. Such an assumption risks bringing with it a bias of a type amply described and analyzed in literature (Weingart, 2004; Moed, 2002; Moed and Van Leeuwen, 1996). In particular, the IF, being an arithmetic mean value of a very skewed distribution of citations, tends to overestimate the real impact of the greater part of the articles that appear in a journal. In the current case, however, since the analysis refers only to publications presented for the VTR, such distortions ought not to be significant: since they are the best outputs to be found in the scientific portfolios of the universities, the VTR selections should receive a number of citations not far from the peak of the citation distributions of their relative journals. Furthermore, it is legitimate to expect that, given the level of aggregation of the measures, the potential distortions would in any case be distributed in a homogenous manner.

---

[12] Mathematics and computer sciences; physics; chemistry; earth sciences; biology; medicine; agricultural and veterinary sciences; civil engineering and architecture; industrial and information engineering; arts and humanities; history, philosophy, pedagogy and psychology; law; economics and statistics; political and social sciences.

[13] Science and technology (ST) for communications and an information society; ST for food quality security, ST for nano-systems and micro-systems; aerospace ST, and ST for the sustainable development and governance.

[14] The ORP registers, for the three-year period 2001 to 2003, the international scientific production of all Italian academic scientists. The main data are based on SCI™, Cd-Rom version: the ORP indexed publications are those containing at least one address corresponding to an Italian university; after which, the ORP applies a disambiguation algorithm to attribute the publication to its respective academic authors. This algorithm is particularly complex given the extremely large number of homonyms in the observation field and the intrinsic limitations of the SCI™ source. These limitations are such that the author list included in the SCI™ repertoire gives the last name but only the *initial* of first name of each author. Furthermore, although the SCI™ lists the authors for each article, alongside a list of the institutions of affiliation, it does not actually indicate which author is specific to which institution.

[15] The normalization procedure is conducted by transforming the absolute values of impact factor into percentile rankings, based on the distribution of the impact factor for all the journals in a given sector. The distribution of the impact factors of journals is actually remarkably different from one sector to another. The operation of normalization makes it possible to contain the distortions inherent in measurements performed over different sectors.

In general, therefore, the average quality of publications from university j in area I ($QI_{ji}$) equals:

$$QI_{ji} = \frac{\sum_{k=1}^{n_{ji}} QI_{kji}}{n_{ji}} \quad [2]$$

$QI_{kji}$ = IF percentile ranking (with respect to SDS distribution) of the journal of article k, of University j, in the disciplinary area i

$n_{ji}$ = number of articles of University j, in the disciplinary area i

The classification of a publication by its individual authors, and not only according to institutional source, further permits quantifying the scientific performance of a scientific-disciplinary sector (SDS) through the aggregation of data referring to all the scientists adhering to that sector[17]. A number of values regarding publication intensity in each of its three dimensions (quantitative, qualitative and fractional) were identified for each area and each university. In particular, the following indicators were used:

- Output (O): total of publications authored by researchers from the university in the survey period;
- Fractional Output (FO): total of the contributions made by the university to the publications, with "contribution" defined as the reciprocal of the number of organizations with which the co-authors are affiliated;
- Scientific Strength (SS): the weighted sum of the publications produced by the researchers of a university, the weights for each publication being equal to the normalized impact factor of the relevant journal;
- Fractional Scientific Strength (FSS): similar to the Fractional Output, but referring to Scientific Strength;
- Productivity indicators (Productivity, P; Fractional Productivity, FP; Qualitative Productivity QP; Fractional Qualitative Productivity, FQP), defined as the ratios between each of the preceding indicators and the number of university staff members in the surveyed period;

---

[16] The source of bibliometric data available to the authors, the SCI™ (CD-rom version) does not contain citation counts.

[17] In the Italian academic system, each scientist is classified as adhering to a scientific disciplinary sector. The 8 disciplinary areas under observation include 183 scientific disciplinary sectors.

Once distributed by sector, the productivity data are re-aggregated by disciplinary area, through the steps shown in [3].

$$P_{ji} = \frac{\sum_{k=1}^{n_i} Pn_{jk} \times Staff_{jk}}{\sum_{k=1}^{n_i} Staff_{jk}} \quad [3]$$

where:
- $P_{ji}$ = the productivity value (P, FP, QP or FQP) of University j in disciplinary area i;
- $Pn_{jk}$ = the productivity value (P, QP, FP or FQP) of University j, within SDS k of disciplinary area i, normalized to the mean of the values of all Universities for SDS k;
- $n_i$ = number of SDS included in disciplinary area i;
- $Staff_{jk}$ = number of staff members of University j affiliated to SDS k of disciplinary area i.

The operations for normalization and weighting of sectorial data make it possible to limit the bias frequently suffered by comparisons performed at high aggregation levels. Different sectors are characterized by different scientific publishing rates and thus robust comparisons are only possible through normalization of the data to the sector average and weighting by the number of personnel in each sector[18]. In this way it is possible to present a productivity rating that is bibliometrically robust, for each university in each disciplinary area, for the period 2001 to 2003 (the same as for the VTR). As an example, Table 3 shows calculations of the Fractional Qualitative Productivity of the University of Padua in the area of mathematics and computer science. During the surveyed period, this university had 121 research scientists distributed over the full 10 sectors of this area. The university registered an FQP score of 0.841, within the total surveyed area and time period.

[Table 3]

## 5. Results of the comparison

### 5.1 Correlation of results from the two evaluation methods

To respond to the first research question, i.e. to verify potential elements of replication and/or complementarities in the results of the peer review and bibliometric methods, the first step was to verify that the results of the VTR were similarly reflected in the bibliometric measures observed by the ORP, in terms of outputs presented by the universities and the outputs assessed by the ORP. Since the VTR evaluation process begins from the submission of an entire range of research outputs by each university (books, book chapters, proceedings of national and international congresses, patents, designs, etc.), while the ORP is based only on listings of publications in international journals, as extracted from the SCI[TM], a preliminary verification was made that the SCI$^{TM}$ data were actually representative, for each area of interest. As shown in Table 4, for the 8 disciplinary areas under observation, the articles in international reviews amount to an average of 95% of the total outputs submitted for the VTR, with a minimum of 89% in Industrial and information engineering and a maximum of 97% in Medicine.

---

[18] For a discussion on this issue, see Abramo and D'Angelo (2007).

[Table 4]

The scope of the analysis then included verification of significant correlation between the two sets of ratings of universities, in each of the 8 areas considered, as issued both from the peer review exercise of the VTR ( $R_{ji}$ ) and from the bibliometric exercise of the ORP ( $QI_{ji}$ )[19].

Table 5 shows results from the analysis: the index of correlation between $R_{ji}$ and $QI_{ji}$ is shown for each area considered, also with the critical value for verification of significance.

[Table 5]

In 6 areas out of 8 the correlation index is greater than 0.6, with a maximum of 0.876 for the agricultural and veterinary sciences area. Only in the industrial engineering and physics areas does the correlation, while still significant, not result so strong. In the first of these areas this result could be due to the relatively low representation of articles among the total of outputs that universities submitted for the VTR. There remains the fact that the correlation is significant in all the areas analyzed and strong in 6 out of 8 of the areas: for these areas, which represent 50% of the entire Italian academic personnel, the results from peer review evaluations are clearly correlated to results observed *a priori* in bibliometric volumes. This demonstrates that both approaches generate similar rankings in measures of quality. These findings from the Italian case confirm those of the British Research Assessment Exercise. The RAE exercise led Oppenheim (1997) to suggest that for reasons of costs, the peer judgments in that exercise should be replaced by citation analyses. Notwithstanding the existence of high correlation between RAE scores and bibliometric measures, according to Warner (2000) there are deviant cases, which should discourage replacing completely peer review with bibliometrics.

Such a result, if not a foregone conclusion, confirms what could logically be expected: an article published in an international journal has already been submitted to peer evaluation, with the peer evaluators being nominated by the editors of the journal itself. Such journals have a selection process for the publications submitted that increases in rigor with increasing number of submissions for publication. It is clear that the most prestigious journals in each sector are also those with the most rigorous selection processes. Such journals typically accept the submissions with greatest originality and innovation, which then in turn receive elevated indexes of citation that bear on the impact factor of the journal.

---

[19] The analysis did not take into consideration the dimensional subdivisions of the universities, done by VTR, since it was not supported by a demonstration of different returns to scale in research activities, in the chosen ranges. Universities that submitted less than 3 outputs for evaluation were also excluded.

**5.2 The effectiveness of processes for selecting research outputs**

Where peer review evaluation exercises like the VTR, which focus on a sampling of the entire research production, include the declared objective of supporting processes for efficient allocation of resources, it is evident that the understood merit criteria is "excellence". The correlation resulting from the comparison of the two approaches (peer review and bibliometric) conducted in the previous paragraph demonstrates the similar value of the two methods in measuring the quality of scientific publications. It is still necessary, however, to verify that individual research institutions have the capacity to effectively their best outputs, which they will then submit for evaluation. If this were not the case, the evaluation exercises based on peer review would produce rankings of quality (and in consequence, resource allocations) that were distorted by an initial ineffective selection.

The selection of outputs within individual research institutions could be influenced by both social and technical factors. Social factors would include the varying negotiating power of persons or groups. For example, negotiations at the outset could favour selection according to the prestige and position of the authors, rather than opting to consider the simple intrinsic quality of outputs (a "quality" selection that then resulted only in the choice of outputs from colleagues would have negative implications concerning the relative value of one's own research). The technical factors mentioned concern the objective difficulty of comparing research outputs from diverse sectors. As an example, one could imagine the difficulty for a university in choosing the best of three hypothetical articles in medical sciences, when each article results from a distinct sector such as dermatology, cardiology or neuroscience. While the problem would not arise in the downward phase (VTR), where the peer reviewers will evaluate all university outputs falling in one single discipline in which they are expert, the problem becomes more difficult to solve at the upward stage (the university level). Support from bibliometric indicators, such as the citations to the article and the impact factor of the journal, could also be misleading if not normalized with respect to the sector averages.

In the authors' view, the selection of research outputs by individual research organizations is the weakest phase in the peer-review process. The aim of the following analysis is thus to verify the validity of the interior selection process of research outputs through the analysis of their positioning, in terms of quality, within the portfolio of publications from each university, in each area. In this regard, the analysis will use the bibliometric variable $QI_{kji}$, previously defined in [2], section 4. Table 6 shows, for each area, the percentage of publications which each university has selected for the VTR, the $QI_{kji}$ is less than the median of the distribution of all the outputs produced by the university in that area.

Such percentages show an average varying from a minimum of 3.7% in biology to a maximum of 29.6% for agricultural and veterinary sciences. Other than this last area, notable figures also emerge for industrial and information engineering (26.5%) and mathematics and computer science (24.8%), as areas in which the selection process results as particularly incoherent. The data seen in the fourth column of Table 6 indicate that in 6 out of 8 areas there were actually universities that selected all articles with a $QI_{kji}$ less than the distribution median for the area.

[Table 6]

As a further detailed illustrative example, Table 7 presents the results of the analyses conducted for the University of Rome "Tor Vergata". In mathematics and computer science, of the 22 outputs presented, only 2 had a value of $QI_{kji}$ that would justify their selection for the VTR. Similar data are also seen in other areas. In chemistry, among the publications authored by the researchers at the university, there were 6 that were not selected for the VTR even though they represented a $QI_{kji}$ that would place them among the top 8 of the entire area. In industrial and information engineering, only 3 of 14 publications presented actually placed among the top 14 publications for the area, as analyzed by $QI_{kji}$.

[Table 7]

It thus emerges that the selection processes adopted by universities for their participation in the VTR are not of a homogenous nature, and instead are found to be rather incoherent with respect to the results of bibliometric measures. Effectively, individual universities did not select their best outputs. Thus, the overall result of the evaluation exercise is in part distorted by an ineffective initial selection, hampering the capacities of the evaluation to present the true level of scientific quality of the institutions.

## 5.3 Quality and productivity

The third and final question that this study intends to address is whether excluding productivity from merit measurement criteria of universities is legitimate. In the authors' view, the assessment of scientific excellence should include the embedded measurement of both quality and productivity. It is to be noted that the VTR classifications capture a quality evaluation based on only 14% of total research production. If rankings by quality evaluation based on the peer review approach agree with rankings of universities based on productivity, it is evident that no conflict occurs. The control that we intend to conduct in the following paragraphs is whether the research institutions evaluated as excellent in terms of quality are also necessarily those that are most efficient in research activities. In particular, the intention is to verify how many among the universities that are at the top in scientific productivity do not appear among the top ranking for quality, according to the VTR. To do this, we will examine each area in every university, comparing the VTR rating ($R_{ji}$) with that referred to the average bibliometric productivity ($P_{ji}$). Table 8 shows the results of the correlation analysis conducted, area by area, for the percentile rankings of the universities resulting from the two approaches[20]. The tabulated values indicate a positive correlation, although not strong for all indicators, in 4 out of 8 areas: the areas of mathematics and computer science, chemistry, medicine, and agricultural and veterinary science. In biology, correlation is completely absent for two indicators and weak for the other two. In physics, earth science, and industrial and information engineering there is a total absence of correlation.

---

[20] In this case again, the analysis excluded universities that submitted less than 3 outputs for evaluation, in order to avoid distortion from outliers.

[Table 8]

Table 9 shows, for each disciplinary area, the statistics describing the variations in ranking variations obtained from the two methods of evaluation: quality (from the VTR) and fractional qualitative productivity (from the bibliometric approach). Such variations concern practically all universities. Agricultural and veterinary sciences seems to be the area with the least variation, yet even for these areas only 4 out of 24 universities retain the same rank, with the universities on average registering a rank change in the order of 4 positions. Conversely, in physics all universities change rank, on average by 16 positions. The maximum changes, in general, present values that are close to the range of variation of the data: in other words, we see that in almost all areas there are universities that result among the top under one type of evaluation and at the bottom according to the other (and vice versa). Finally, the comparison of the top 10 universities by disciplinary area in the two rankings (corresponding to approximately 20% to 25% of the total number of universities) shows that only a small portion (less than 30%) of those that result at the top in quality (VTR) also excel in fractional qualitative productivity (ORP). In physics, only one university results in the top 10% in both rankings; in biology there are 2; in mathematics, medicine, and industrial and information engineering, there are only 3. From the point of view of funding allocations, this means that, on average, 7 of 10 of the top (in terms of qualitative productivity) universities would receive less money than those that rank still lower on the scale.

The results of the analysis thus seem to indicate that the peer review approach (based on a sampling of outputs, and thus only intended to evaluate the level of quality of institutions), provides rankings that are quite different from those referred to research qualitative productivity, measured as the ratio of outputs (in this case, the quantity and quality of publications) and resources employed (numbers of scientific personnel on staff). The research institutions rated as top in quality of production do not necessarily also rank well in research qualitative efficiency.

[Table 9]

## 6. Conclusions

The obligation to improve the research productivity of public institutions and provide them with resources in an efficient manner plays a leading role in the political agendas of increasing numbers of developed nations. In such a context, it is imperative to define and implement effective evaluation systems that, in support of the allocation processes, stimulate adoption of strong strategy and practices for increased productivity, both in quality and quantity, by universities and public research organizations.

Comparative evaluations of institutions in public research systems are, however, not exercises without risk of pitfalls. In particular, outcome control types of approaches risk ignoring the strategic dimensions of the allocation problem. It can be argued that such evaluations result in allocation of limited resources according to more or less fixed concepts of "excellence", measured through universal, static algorithms and methods, irrespective of the external and internal context of research institutions, instead of following a more articulated series of strategic criteria (Abramo, 2006). Nevertheless, outcome control types of evaluation exercises are widely diffused, and often guide allocation choices for notable segments of national budgets destined for research. The peer review method is the most widely used for this purpose, while bibliometric methods are, at the most, employed to provide integrative data to support systems based on peer review.

The recent publication of results from the first national exercise in evaluation carried out for the Italian academic system (the VTR) provides an important opportunity to compare the peer review approach (as adopted by the VTR) with the bibliometric approach (based on the Observatory of Public Research for Italy, developed by the authors), the objective being to investigate to what extent the relatively inexpensive, time saving and exhaustive bibliometric methodology is a satisfactory and appropriate replacement or enhancer for peer review.

In the hard sciences, comparative analysis of the ratings obtained from the VTR and from proxy bibliometric measures of the average quality of the set of articles under VTR evaluation reveal a significant overlap in the results of the two approaches. For evaluation of the quality of research outputs (of a specific institution, in a given disciplinary area) the two methods are thus substantially equivalent. There is, after all, no reason to believe that an evaluation of an article's quality by two experts nominated as part of a national evaluation exercise would be better than the evaluation by international referees on behalf of the journal in which it is published and of the peers who then cite the article. The differences in cost and times to execute the evaluations would certainly be relevant; however, the underlying situation causing concern is the process by which each university selects the articles to be submitted to the national evaluation committee. Both social and technical factors (parochialism; the real difficulty of comparing articles from diverse disciplines) can clearly compromise the effectiveness of the selection process.

This study then also addressed the question of the appropriateness of the approaches with respect to the objective for which they are conceived. An evaluation exercise based on peer review typically focuses on a sampling of the totality of scientific outputs, with the aim, then, of evaluating their relative quality. The above-noted limits during the phase of selecting outputs to be submitted for evaluation can, however, bring about distorted ratings for quality and a consequent inefficient allocation of resources, where that has been the purpose of the evaluation. Indeed, bibliometric analysis of the publications selected from the hard science areas for participation in the VTR shows that Italian universities, in the main, did not identify and/or present their supposedly best publications. It is thus legitimate to doubt that the ratings obtained could reflect the real level of quality of the organizations rated, or that allocation of resources based on such evaluation could be efficient.

What is more, basing a merit judgment and the consequent allocation of resources on the qualitative evaluation of a very limited sample of the total research output, completely ignoring measurement of the productivity, could also raise serious doubts. The investigations conducted in the third part of this study indeed demonstrate that the universities indicated as being of top quality by the VTR are not necessarily also those that are most productive, in both quantitative and qualitative terms. As a direct cause of this, more productive research institutions would receive fewer funds than would be economically optimal, from the prospective of allocative efficiency.

The authors are distinctly aware that in the arts and humanities, as in law and in part in socio-economic areas, international scientific publication does not represent the usual form for codifying and disseminating the results of one's research activity, and the peer review approach thus remains difficult to substitute. However, for the hard sciences (which in a country such as Italy represent almost two thirds of the entire academic system), and having overcome the problem of the accurate attribution of articles to their true authors and institutions, bibliometrics currently offer levels of potential and methodological maturity that should induce a reconsideration and revision of their role. Comparing to the peer review process, the bibliometric approach permits:
- i) avoiding the weakest phase in peer review, meaning the selection of articles by individual research institutions;
- ii) assessing research productivity, both in quantitative and qualitative terms;
- iii) significantly reducing the costs and times for implementation.

The peer-review approach still remains necessary to evaluate research outputs other than publications in indexed journals (such as patents, proceedings, etc.) and dimensions of excellence other than quality and productivity, particularly the dimension of the socio-economic impact of research activities.

| University | Selected outputs | E | G | A | L | Rating | Category rank | Absolute rank | Absolute rank (percentile) |
|---|---|---|---|---|---|---|---|---|---|
| Milano | 28 | 17 | 10 | 1 | 0 | 0.914 | 1 | 4 | 92 |
| Milano Politechnic | 25 | 16 | 7 | 2 | 0 | 0.912 | 2 | 6 | 90 |
| Pisa | 42 | 22 | 18 | 2 | 0 | 0.895 | 3 | 9 | 85 |
| Roma "La Sapienza" | 61 | 31 | 26 | 4 | 0 | 0.889 | 4 | 13 | 77 |
| Bologna | 35 | 17 | 15 | 3 | 0 | 0.880 | 5 | 16 | 67 |
| Padova | 31 | 11 | 17 | 3 | 0 | 0.852 | 6 | 23 | 58 |
| Firenze | 31 | 12 | 15 | 3 | 1 | 0.839 | 7 | 25 | 54 |
| Palermo | 31 | 9 | 14 | 7 | 1 | 0.794 | 8 | 39 | 27 |
| Torino | 30 | 7 | 15 | 7 | 1 | 0.780 | 9 | 41 | 19 |
| Genova | 30 | 7 | 17 | 4 | 2 | 0.780 | 9 | 41 | 19 |
| Napoli "Federico II" | 43 | 7 | 26 | 8 | 2 | 0.767 | 11 | 44 | 17 |

*Table 1: VTR rank list of Italian universities, of category "large", for the disciplinary area mathematics and computer science; "E, G, A, L" indicate number of outputs rated as "excellent, good, acceptable, limited"*

| Disciplinary area | Selected outputs | E | G | A | L | Rating | Category rank |
|---|---|---|---|---|---|---|---|
| Mathematics and computer science | 23 | 17 | 5 | 1 | 0 | 0.939 | 1 out of 15 (medium) |
| Physics | 19 | 10 | 9 | 0 | 0 | 0.905 | 8 out of 23 (medium) |
| Chemistry | 8 | 3 | 5 | 0 | 0 | 0.875 | 7 out of 26 (small) |
| Biology | 38 | 21 | 13 | 4 | 0 | 0.889 | 5 out of 23 (large) |
| Medicine | 93 | 23 | 51 | 12 | 7 | 0.778 | 10 out of 16 (very large) |
| Civil engineering and architecture | 10 | 2 | 5 | 3 | 0 | 0.780 | 5 out of 15 (medium) |
| Industrial and information engineering | 21 | 5 | 10 | 2 | 4 | 0.714 | 18 out of 18 (medium) |
| Arts and humanities | 23 | 13 | 6 | 3 | 1 | 0.861 | 12 out of 17 (medium) |
| History, philosophy, pedagogy and psychology | 15 | 6 | 7 | 2 | 0 | 0.853 | 2 out of 15 (medium) |
| Law | 28 | 4 | 17 | 5 | 2 | 0.750 | 9 out of 15 (large) |
| Economics and statistics | 18 | 0 | 7 | 4 | 7 | 0.522 | 28 out of 31 (medium) |

*Table 2: VTR ratings for University of Rome "Tor Vergata"; E, G, A, L indicate number of outputs rated as "excellent, good, acceptable, limited"*

| SDS | Staff | O | FSS | FQP | FQP (Average all Univ.) | FQP (normalized) | FQP (norm. and weighed) |
|---|---|---|---|---|---|---|---|
| MAT/01 - Mathematical logic | 3 | 5 | 0.789 | 0.237 | 0.144 | 1.644 | 0.041 |
| MAT/02 - Algebra | 14 | 27 | 6.000 | 0.419 | 0.207 | 2.022 | 0.234 |
| MAT/03 - Geometry | 22 | 21 | 1.718 | 0.078 | 0.159 | 0.491 | 0.089 |
| MAT/04 - Applied mathematics | 3 | 0 | 0 | 0 | 0.214 | 0 | 0 |
| MAT/05 - Mathematical analysis | 35 | 28 | 4.546 | 0.131 | 0.283 | 0.463 | 0.134 |
| MAT/06 - Statistics & probability | 7 | 5 | 0.926 | 0.132 | 0.286 | 0.462 | 0.027 |
| MAT/07 - Mathematical physics, | 13 | 15 | 5.929 | 0.456 | 0.660 | 0.691 | 0.074 |
| MAT/08 - Numerical Analysis | 9 | 21 | 5.947 | 0.686 | 0.302 | 2.272 | 0.169 |
| MAT/09 - Operations Research | 8 | 14 | 1.137 | 0.136 | 0.318 | 0.429 | 0.028 |
| INF/01 – Computer science | 7 | 8 | 1.819 | 0.260 | 0.334 | 0.778 | 0.045 |
| Total | 121 | | | | | | 0.841 |

**Table 3: Bibliometric performance of each scientific disciplinary sector within the disciplinary area "mathematics and computer science", for the University of Padua (average values during 2001-2003 period)**

| Disciplinary area | Selected outputs | Of which "articles" | Total research staff (average 2001-03) |
|---|---|---|---|
| Medicine | 2,644 | 2,574 (97.4%) | 10,340 (18.9%) |
| Biology | 1,279 | 1,239 (96.9%) | 4,806 (8.8%) |
| Industrial and information engineering | 909 | 807 (88.8%) | 4,334 (7.9%) |
| Chemistry | 758 | 712 (93.9%) | 3,132 (5.7%) |
| Mathematics and computer science | 751 | 711 (94.7%) | 3,030 (5.5%) |
| Physics | 626 | 596 (95.2%) | 2,502 (4.6%) |
| Agricultural and veterinary science | 617 | 571 (92.5%) | 2,939 (5.4%) |
| Earth science | 323 | 303 (93.8%) | 1,276 (2.3%) |
| *Sub tot.* | *7,907* | *7,513 (95.0%)* | *32,359 (59.1%)* |
| Arts and humanities | 1,278 | 103 (8.1%) | 5,410 (9.9%) |
| History, philosophy, pedagogy and psychology | 1,123 | 249 (22.2%) | 4,376 (8.0%) |
| Law | 1,019 | 140 (13.7%) | 3,964 (7.2%) |
| Economics and statistics | 953 | 691 (72.5%) | 3,807 (6.9%) |
| Civil engineering and architecture | 752 | 398 (52.9%) | 3,495 (6.4%) |
| Political and social science | 342 | 66 (19.3%) | 1,371 (2.5%) |
| *Tot.* | *13,374* | *9,160 (68.5%)* | *54,786* |

*Table 4: Representation of scientific articles among outputs selected for the VTR, with totals of academic research staff (data 2001-2003, for Italian universities, for each disciplinary area)*

| Disciplinary area | Correlation index | Critical value (at 5%, two tails) |
|---|---|---|
| Agricultural and veterinary science | 0.876 | 0.404 (for 24 observations) |
| Biology | 0.743 | 0.279 (for 50 observations) |
| Earth science | 0.668 | 0.339 (for 34 observations) |
| Chemistry | 0.645 | 0.312 (for 40 observations) |
| Medicine | 0.627 | 0.316 (for 39 observations) |
| Mathematics and computer science | 0.641 | 0.271 (for 47 observations) |
| Physics | 0.409 | 0.297 (for 44 observations) |
| Industrial and information engineering | 0.336 | 0.312 (for 40 observations) |

*Table 5: Correlation analysis between ratings of universities deriving from the VTR peer review ( $R_{ji}$ ) and the bibliometric ORP ( $QI_{ji}$ ), for each disciplinary area*

| Disciplinary area | Average | Median | Max | Variation coefficient |
|---|---|---|---|---|
| Agricultural and veterinary science | 29.6% | 26.3% | 100% | 0.912 |
| Industrial and information engineering | 26.5% | 26.0% | 100% | 0.868 |
| Mathematics and computer science | 24.8% | 24.0% | 100% | 0.897 |
| Earth science | 17.4% | 14.3% | 100% | 1.179 |
| Physics | 8.5% | 0% | 100% | 1.939 |
| Chemistry | 5.0% | 0% | 100% | 3.312 |
| Medicine | 3.8% | 1.2% | 33.3% | 1.895 |
| Biology | 3.7% | 0% | 35.3% | 2.105 |

*Table 6: Statistics for percentage of outputs selected in each area with quality level ( $QI_{kji}$ ) that is less than the median of the total university publication portfolio in the area*

| Disciplinary area | Articles selected for VTR | Articles in ORP | Articles with $QI_{kji}$ greater than those selected for VTR (%) |
|---|---|---|---|
| Mathematics and computer science | 22 | 242 | 20 (90.9%) |
| Industrial and information engineering | 14 | 289 | 11 (78.6%) |
| Chemistry | 8 | 296 | 6 (75.0%) |
| Physics | 17 | 497 | 11 (64.7%) |
| Medicine | 92 | 996 | 59 (64.1%) |
| Biology | 38 | 440 | 20 (52.6%) |

*Table 7: For the University of Rome "Tor Vergata": analysis of positioning of outputs selected for the VTR within the total publication portfolio of each disciplinary area*

| Disciplinary area | P | FP | QP | FQP | Critical value (at 5%, two tails) |
|---|---|---|---|---|---|
| Medicine | 0.593 | 0.623 | 0.556 | 0.615 | 0.316 (for 39 observations) |
| Agricultural and veterinary science | 0.628 | 0.443 | 0.691 | 0.532 | 0.404 (for 24 observations) |
| Mathematics and computer science | 0.440 | 0.451 | 0.501 | 0.498 | 0.271 (for 47 observations) |
| Chemistry | 0.503 | 0.399 | 0.566 | 0.489 | 0.312 (for 40 observations) |
| Biology | 0.301 | 0.131 | 0.391 | 0.230 | 0.279 (for 50 observations) |
| Industrial and information engineering | 0.242 | 0.220 | 0.250 | 0.226 | 0.312 (for 40 observations) |
| Earth science | 0.249 | 0.138 | 0.260 | 0.151 | 0.339 (for 34 observations) |
| Physics | 0.095 | -0.067 | 0.074 | -0.056 | 0.297 (for 44 observations) |

*Table 8: Correlation analysis, by area, of VTR percentile ranking (index $R_{ji}$) and percentile productivity rankings (index $P_{ji}$) from four bibliometric indicators*

| Disciplinary area | Variations | Max | Average | Median | Standard Deviation | Universities in both top 10 rankings |
|---|---|---|---|---|---|---|
| Physics | 52 of 52 (100%) | 51 | 16..5 | 15.5 | 12.0 | 1 |
| Earth sciences | 34 of 34 (100%) | 30 | 9.5 | 7 | 7.7 | 4 |
| Industrial and information engineering | 39 of 40 (97.5%) | 34 | 11.4 | 9 | 9.0 | 3 |
| Chemistry | 38 of 39 (97.4%) | 27 | 8.0 | 6 | 6.7 | 6 |
| Medicine | 37 of 39 (94.9%) | 24 | 8.5 | 7 | 7.1 | 3 |
| Biology | 45 of 48 (93.8%) | 37 | 12.5 | 11 | 10.1 | 2 |
| Mathematics and computer sciences | 43 of 46 (93.5%) | 33 | 10.1 | 8 | 8.6 | 3 |
| Agricultural and veterinary sciences | 20 of 24 (83.3%) | 13 | 4.1 | 4 | 3.6 | 7 |

*Table 9: Ranking variations of universities in VTR scores compared to FQP scores for each disciplinary area*